# Polychromatic angle resolved IBIC analysis of silicon power diodes.

M. Pezzarossa[1], E. Cepparrone[2], D. Cosic[4], M. Jakšić[4], G. Provatas[4], M. Vićentijević[4] , E. Vittone[2,3]*

[1] Department of Applied Science and Technology (DISAT), Politecnico di Torino (Italy)

[2] Department of Physics, NIS interdepartmental center, University of Torino, INFN, Torino Unit, via P. Giuria 1, 10125 Torino, Italy.

[3] INFN Torino unit, via P. Giuria 1, 10125 Torino, Italy.

[4] Department for Experimental Physics, Ruđer Bošković Institute, Bijenička cesta 54, 10000 Zagreb, Croatia.

**Abstract**

This paper describes both an experimental methodology based on the Ion Beam Induced Charge (IBIC) technique and the relevant interpretative model, which were adopted to characterize the electronic features of power diodes.

IBIC spectra were acquired using different proton energies (from 1.2 to 2.0 MeV), angles of incidence, and applied bias voltages.

The modulation of the ion probe range, combined with the modulation of the extensions of the depletion layer, allowed the charge collection efficiency scale to be accurately calibrated, the dead layer beneath the thick (6 μm) Al electrode and the minority carrier lifetime to be measured.

The analysis was performed by using a simplified model extracted from the basic IBIC theory, which proved to be suitable to interpret the behaviour of the IBIC spectra as a function of all the experimental conditions and to characterize the devices, both for what concerns the electrostatics and the recombination processes.

## 1. Introduction

* Corresponding author: ettore.vittone@unito.it

Power electronics is the key technology to control the flow of electricity from the source to the load and is the backbone of the whole power supply infrastructure in our society from the energy generation, transmission, distribution and, for its pervasiveness in a huge variety of applications, it plays a fundamental role in energy saving and improved energy efficiency [1].

Power electronics is mainly based on power semiconductor devices [2]. Their main attractiveness in this sector derives from their ability to fast switching from the "on-" to "off-" state, where "on-state" corresponds to the condition in which the current can ideally flow without losses and "off-state" the condition in which the current is ideally blocked, without leakages.

Among the many key parameters, which influence device performances, carrier lifetime plays a dominant role due to its impact on the determination of the reverse recovery time, i.e. the duration of the recovery transient from the highly conductive to the blocked states, which controls the efficiency for power conversion [3] [4].

It is then unavoidable to develop and to apply experimental methods and models, possibly with non-invasive or non-destructive approaches, to assess the efficiency of processes adopted to control the carrier lifetimes [5][6].

In the literature, the several methods that have been proposed to reach this goal are essentially based on the measurement of effects related to the decay of excess carriers generated in the neutral region.

The injection of excess carriers is usually performed by electrical or optical means [7][8]. For the former, carriers are injected by the heavily doped region and the open-circuit voltage or the diode current transient measurements allow the extraction of carrier lifetimes. However, the reliability of these measurements, which is bound to an accurate knowledge of the diode structure, of the surface and of the heavily doped region recombination lifetimes, was proven to be inadequate for the Si power diode characterization [9],

Optical methods are extensively used for photonic [10] and wide bandgap [11] devices, offering the advantage of a local characterization, by scanning the focused laser beam. However, these techniques are limited to the study of the periphery of the active junctions or devices with transparent electrodes, rarely encountered in the field of power semiconductor devices.

The imaging of buried sub-electrode active region can be carried out by the electron beam induced current technique (EBIC), which relies on the use of focused electron beams with energies from few to tens keV [12] [13]. However, especially in power devices, Al pads of the order of a few μm thick are usually adopted, to reduce the impact of bonding wires and prevent the damage of the underlying silicon [14][15]. Such thicknesses hinder the penetration of the electron beam probes and/or make difficult the estimate of the effective electron/hole generation rate and volume into the active semiconductor regions beneath the electrode and prevent the use of photon or electron probes to characterize the bulk properties of the diode.

The Ion Beam Induced Charge (IBIC) technique provides measurements of basic parameters of the active region under metallisation layers. The IBIC technique uses MeV light ion microbeams, focused to spot sizes of the order of micrometer scanned over the electrode. The ion current is very low, of the order of fA (thousands of ions/s), in order to avoid any radiation damage and to allow individual charge pulses to be detected by virtue of the high number of electron/hole pairs (EHPs), generated within few tens of nanometers from the ion track. The major aspects of the IBIC technique are covered in Ref. [16].

For example, 2 MeV protons, with a longitudinal and projected range in silicon of 48 μm and 1.5 μm, respectively [17], generate around $5 \cdot 10^5$ EHPs along the ion trajectory. Therefore, single ions induce measurable charge pulses above the noise level and the active regions of the device, which can be buried so deep under thick electrodes, can be analysed with a micrometer lateral resolution, which is reasonable for the study of many semiconductor power devices.

Finally, a further advantage offered by the IBIC technique stems from the availability of a rigorous theoretical formalism [18], which is able to model the charge pulse formation and allow the extraction of basic transport and recombination parameters, which are essential for a reliable design of the device.

IBIC technique has already been applied to many studies of semiconductor materials and devices [16]. In particular, in the field of high power electronics, thyristors have been studied by Osipovitz, Smeck et al. [19] [20] and power diodes with lifetime killers induced by He irradiation have been studied by Fizzotti et al.[21].

The former investigation was non-destructive and provided CCE images, which were qualitatively related to the electric field distribution in deeply buried junctions, whereas the latter research was able to measure the carrier diffusion length by means of the lateral IBIC approach, which is notably destructive.

In this paper, we report on a methodology based on the IBIC technique, able to non-invasively and quantitatively investigate the features of silicon power diodes. Proton microbeams of energy ranging from 1.2 to 2.0 MeV, incident onto the diode top electrode at different angles, were used to measure induced charge spectra at different bias voltages.

The use of different ion energies, allows a fine probing of the active region, ranging from the junction to the bulk of the intrinsic region and provide experimental data, which need to be consistently analysed for all the experimental parameters (tilting angles, ion energies, applied bias voltages). Therefore, this approach requires an accurate interpretative model, derived from the basic IBIC theory, which overcomes the inherent limitations of that previously adopted [22], and a proper data analysis, suitable to consistently fit all the experimental data.

## 2. Materials and methods

The analysis was carried out on two commercially available p-i-n power diodes fabricated on a 60 μm thick epitaxial silicon layer grown onto a Czochralski $n^+$ substrate (Fig. 1a).

Diode type 1 is an ordinary rectifier diode, whereas diode type 2 is a fast recovery diode (FRD) [7]. The two devices show the same structure but differ in their switching features: the reverse recovery time scale is of the order of few microseconds for the former and tens/hundreds of nanoseconds for the latter because of the adoption of lifetime killing technologies.

Each diode was mounted on a PCB; the (2.5x2.5) $mm^2$ back contact (cathode) was directly fixed to a copper pad by conductive glue, and the (1.3x1.3) $mm^2$ Al top electrode (6 μm thick) was connected to a different Cu pad by 20 μm Al wire (Fig. 1b). The floating field ring structure, surrounding the central electrode, covers an area of (1.7x1.7) $mm^2$.

The PCB was mounted onto a rotating holder connected to a goniometer allowing the orientation of the sample with respect to the incident ion beam axis with an uncertainty of 1°.

Although the device can operate up to reverse bias voltages higher than 1000 V, in this study the maximum applied bias was 200 V. As shown in Fig. 1c, for applied bias voltages higher than 100 V, the C-V curve is rather flat, indicating that the extension of the depletion layer saturates.

The same figure shows the $1/C^2$ vs. V plot, whose almost linear behaviour in the first part of the graph, allows the estimation of the effective donor concentration in the $n^-$ region to be of the order of $10^{13}$ $cm^{-3}$. The two diodes present very similar C-V characteristics, but different current-voltage curves, as shown in Fig. 1d. This different current behaviour is to be attributed to the impact in the type 2 diode of impurities (as Au or Pt) or defects generated by high energy particle (e.g. electron) irradiation, which act as lifetime killers to improve the removal of minority carriers by recombination processes and generally to increase the turn-off speed in high-frequency applications [2][4].

At 200 V, the leakage current is below 200 nA, which generates a reasonably low noise in the charge sensitive preamplifier. Similarly, at low bias voltage, the capacitance is less than 50 pF, which falls in the range of capacitance of typical silicon nuclear detectors. These features allow the use of a standard electronic chain used in nuclear radiation spectroscopy to process the induced charge signals. In this work, the electronic chain includes a charge sensitive preamplifier (ORTEC 142a), and spectroscopy amplifier (ORTEC 570). The effective bias voltage across the device was evaluated by subtracting the voltage drop across the 100 MΩ load resistor of the preamplifier from the externally applied bias voltage. In all the measurements the shaping time of the amplifier was set to 2 μs.

The experiments were performed at the accelerator facility of the Laboratory for Ion Beam Interactions of the Ruder Boskovic Institute in Zagreb, Croatia. The ion microbeam line was coupled to the 1.0 MeV Tandetron accelerator [23], while the data acquisition and ion micro-beam scanning system was controlled by the SPECTOR hardware / software package [24] .

Protons with energies of 1.2, 1.5. 1.7 and 2.0 MeV were focused down to a spot size of less than 5 μm and the rarefied (few hundreds of ions per second) proton beams were scanned over the silicon diode frontal electrode.

Fig.2 shows the irradiation set-up and the colour coded pulse-height IBIC maps resulting from the scan of proton microbeams at different tilting angles (θ) (Figs. 2a/2b), proton energies (E) (Figs. 2a/2c) bias voltages (V) (Figs 2c/2d). Since IBIC maps show a rather uniform response, the following analyses were carried out on spectra with typically few thousands of pulses emerging from the irradiation of small, about 0.1 $mm^2$, regions on the frontal electrode. For each spectrum, the proton fluence was less than $10^7$ protons/$cm^2$, which ensures negligible radiation damage effects induced by the ion probe [18].

## 3. Calibration and dead layer measurement

IBIC experiments require an accurate calibration of the electronic chain, in order to convert the induced signal into the charge collection efficiency (CCE), which is defined by the ratio of the induced charge measured at the sensing electrode on the total charge generated by ionization.

The plot of the peak channels, evaluated through a Gaussian fit of the experimental data (the uncertainty in the peak position is around 1 channel), as a function of bias voltages, shows a clear saturation of the curves for V>100 V, independently from the proton energies, (Fig. 3). In the same figure, the depth of the depletion layer w (see Fig. 1) is shown on the top axis in correspondence of the applied bias (bottom axis); vertical lines represent the ranges of protons in the Al/Si diode, simulated by the SRIM2013 software [17], at different energies. A full charge collection occurs at the sensing electrode induced by all the EHPs generated by ionization within the depletion region, since the drift time is short enough to make negligible any carrier recombination process [16][18].

However, the energy deposited by ionization within the active region does not correspond to the proton energy, since a non-negligible fraction of energy is lost in the thick (6 μm) Al electrode and in the silicon dead-layer (or entrance window), i.e. the silicon layer beneath the electrode in which charge collection is inefficient [25].

To evaluate the effective thickness of the entrance window, and to calibrate the electronic chain, we applied angle-resolved IBIC analysis [22], which consists in tilting the sample at angles (θ) ranging from -40° to 50° with respect to the ion beam axis at a bias voltage of 200 V.

Fig. 4a shows the peak channels vs. the tilt angle per different proton energies at a bias voltage of 200 V; the diode response decreases as the tilting angle θ increases due to the increase in ion path length in the non-active regions, actually the Al electrode of thickness t and the dead layer in silicon of thickness d.

The energy $E_{ij}$ deposited in the active region from the proton with energy $E_i^0$ incident at an angle $\theta_j$ is then

(1) $$E_{ij} = E_i^0 - \Sigma_{i,j}^{Al} - \Sigma_{i,j}^{Si} = \epsilon_{ij} - \Sigma_{i,j}^{Si}$$

Where

(2) $$\Sigma_{i,j}^{Al} = \int_0^{\frac{t}{\cos(\theta_j+\theta_{0i})}} \left[\frac{dE_i}{dx}\right]_{Al}$$

is the energy loss in the Al electrode by the proton,

(3) $$\epsilon_{ij} = E_i^0 - \Sigma_{i,j}^{Al}$$

is the proton energy emerging from the Al electrode and

(4) $$\Sigma_{i,j}^{Si} = \int_0^{\frac{d}{\cos(\theta_j+\theta_{0i})}} \left[\frac{d\epsilon_{ij}}{dx}\right]_{Si} \approx \frac{\left[\frac{d\epsilon_{ij}}{dx}\right]_{Si}}{\cos(\theta_j+\theta_{0i})} d$$

is the energy loss in the silicon dead layer by the proton incident at an angle $\theta_j$ and emerging from the Al electrode with an energy $\epsilon_{ij}$ . Assuming the dead layer so thin, that the stopping power is approximately constant, $\Sigma_{i,j}^{Si}$ can be approximated by the last term in eq. 4.

$\theta_{0i}$ is the angular offset, which was evaluated through the analysis of data in Fig. 4a for each energy. Actually, both diodes were measured at a fixed energy, but, since the sample holder can host a single PCB, the angular offset $\theta_{0i}$ was evaluated for each energy $E_i^0$, to take into account the misalignment in the mounting of the PCB onto the goniometer. The angular offsets (of the order of 1°) were evaluated by fitting, for each proton energy $E_i^0$ , the experimental data in Fig. 4a through the standard formula [25]

(5) $$f(\theta) = A_i - \frac{B_i}{\cos(\theta+\theta_{0i})}$$

Assuming a linear relationship between pulse heights (channels Ch) and the energy $E_{ij}$ deposited in the active region:

(6) $E_{ij} = \alpha + \beta \cdot Ch_{ij}$

Where $\alpha$ and $\beta$ are the calibration parameters and $Ch_{ij}$ is the peak position (channels) relevant to charge pulses induced by protons with energy $E_i^0$ incident at an angle $\theta_j$.

From eqs. 0 and 0, we can easily get the following linear expression:

(7) $\epsilon_{ij} = \alpha + \beta \cdot Ch_{ij} + d \cdot \mu_{ij}$

where $\mu_{ij} = \left[\frac{d\epsilon_{ij}}{dx}\right]_{Si} \frac{1}{\cos(\theta_j+\theta_{0i})}$ . The uncertainty ($\Delta\epsilon_{ij}$) of the dependent variables $\epsilon_{ij}$ is evaluated assuming an uncertainty of 100 nm on the electrode thickness; all these terms have been calculated through SRIM software [17].

The calibration parameters ($\alpha, \beta$) and of the silicon dead layer thickness (d) have been extracted from the overdetermined system of linear L=MxN equations (where M=4 is the number of proton energies and N =5 is the number of the tilting angles $\theta_j$ values) given by eq. 7) using the method of least squares [26], where the parameter vector $\vec{X} = \begin{pmatrix} \alpha \\ \beta \\ d \end{pmatrix}$ is linearly connected to the dependent variable column vector $\vec{Y}$ of dimension L through the (Lx3) matrix [A]:

$$\vec{Y} = \begin{pmatrix} \epsilon_{11} \\ \epsilon_{12} \\ \cdot\cdot \\ \epsilon_{21} \\ \cdot\cdot \\ \epsilon_{44} \\ \epsilon_{45} \end{pmatrix} = [A] \cdot \vec{X} = \begin{pmatrix} 1 & Ch_{11} & \mu_{11} \\ 1 & Ch_{12} & \mu_{12} \\ \cdot\cdot & \cdot\cdot & \cdot\cdot \\ 1 & Ch_{21} & \mu_{21} \\ \cdot\cdot & \cdot\cdot & \cdot\cdot \\ 1 & Ch_{44} & \mu_{44} \\ 1 & Ch_{45} & \mu_{45} \end{pmatrix} \cdot \begin{pmatrix} \alpha \\ \beta \\ d \end{pmatrix}$$

and the covariance matrix [W] of $\vec{Y}$ is given by the LxL diagonal matrix

$$[W] = \begin{pmatrix} (\Delta\epsilon_{11})^2 & 0 & \cdot\cdot & 0 & \cdot\cdot & 0 & 0 \\ 0 & (\Delta\epsilon_{12})^2 & \cdot\cdot & : & \cdot\cdot & : & : \\ : & 0 & \cdots & 0 & \cdot\cdot & : & : \\ : & : & \cdot\cdot & (\Delta\epsilon_{21})^2 & \cdot\cdot & : & : \\ : & : & \cdot\cdot & 0 & \cdot\cdot & 0 & : \\ : & : & \cdot\cdot & : & \cdot\cdot & (\Delta\epsilon_{44})^2 & 0 \\ 0 & : & \cdot\cdot & 0 & \cdot\cdot & 0 & (\Delta\epsilon_{45})^2 \end{pmatrix}.$$

The result of this fitting procedure applied to both the diodes is summarized in Fig. 4b and provides the dead layer thickness d=(0.3±0.1) μm; the sensitivity of the electronic ($\beta$) chain is (2.311 ± 0.007) keV/channel corresponding to about 640 electrons/channel as evaluated assuming an average energy to create one electron/hole pair ($\varepsilon_{eh}$) in silicon of 3.6 eV [27].

Such a calibration has been adopted for all the measurements carried out for any point in the experimental parameter space (E,θ,V). The FWHM of the peaks (e.g.: inset in Fig. 4a), considering all the bias voltages, ranges from a minimum of 37 keV (≈10000 electrons) to a maximum of 52 keV (14400 electrons), for (E,θ,V)=(2.0 MeV,0°,200 V) and (E,θ,V)=(1.5 MeV,50°,1 V), respectively.

It is worth mentioning that this procedure allows a consistent calibration of the electronic chain using all the ion energies without any reference detector and is particularly suited for devices with thick electrodes (by virtue of eq. 2), where the only basic linear relationship (eq. 6) cannot be adopted, due to the not negligible energy loss outside the active region.

## 4. CCE profiles and carrier lifetime measurements

To investigate the bulk electronic properties of the diodes, we exploited the combination of the tilting angle θ and of the ion energy, which provides ionization profiles with Bragg's peaks ranging from 6 to 41 μm below the silicon surface (see Fig. 5).

Taking into account that IBIC signals derive from the charge induced by the motion of free carriers into the depletion region [18][28], the depth profiling methodology based on varying the ion beam energy (E) and incident angle (θ), successfully applied for differential PIXE studies of art objects [29], is further potentiated by the control of the depletion region depth as a function of the applied bias voltage (V).

The (few hundreds) IBIC spectra acquired by modulating the three independent parameters, namely (E, θ, V), can be studied by assuming constant one parameter and analysing the behaviour of CCE vs. the other two. Here, we choose an approach in which the CCE is analysed as a function of the incident angle, at a constant applied bias voltage, for different ion energies, as shown in Fig. 6 for the type 2 diode.

It is apparent that the CCE vs. θ behaviours show a monotone decreasing trend at high bias voltages (>50 V) as already observed in Fig. 4a, referring to 200 V bias voltage.

When the extension of the depletion layer is beyond the range of the ions, all the carriers generated by ionization rapidly drift toward the electrodes, and the induced charge can be reasonably considered fully collected. The different decreasing trends of the curves is due to the dependence of the energy loss in the electrode as a function of the proton energy: the lowest ion energy (1200 keV) shows the most pronounced decrease due to the highest stopping power in Al with respect to that for higher protons energies and then a higher sensitivity to the electrode and dead layer thicknesses.

On the other hand, at bias voltages lower than 50 V, the CCE curves in Fig. 6 show seemingly unpredictable shapes.

However, a simplified formulation of the IBIC theory [18][28] can provide a satisfactory interpretation of these behaviours using the following equation:

(8) $$\mathrm{CCE}_{\mathrm{model}}\left(E_i^0, \theta_j, V_l\right) = \frac{1}{E_i^0}\int_0^D \eta(x; V_l)\cdot\left[\frac{dE}{dx}\right]_{E_i^0,\theta_j} dx$$

where $\left[\frac{dE}{dx}\right]_{E_i^0,\theta_j}$ is the stopping power profile shown in Fig. 5 as a function of the incidence angle and the proton energy, D identify the end of the active region (Fig. 1a) and $\eta(x; V_l)$ is the spatial profile of the charge collection efficiency at the applied voltage $V_l$. Such a profile can be modelled as null for the electrode and dead layer thicknesses (i.e. for x<t+d) and one in the depletion region (i.e. for t+d<x<w($V_l$)), where it is assumed that the high drift velocity of the charge carriers under the influence of the electrical field leads to a fast charge collection process compared to the lifetime of the charge carriers.

In the neutral region (D>x> w($V_l$)), only the fraction of minority carriers (holes) reaching the edge of the depletion region contribute to the total induced charge. In this region, assuming constant transport and recombination properties within the neutral region, from the basic IBIC theory [16][18][28], $\eta(x; V_l)$ can be evaluated by solving the equation

(9) $$\frac{d^2\eta(x;V_l)}{dx^2} = \frac{\eta(x;V_l)}{{L_h}^2}$$

where $L_h$ is the hole diffusion length.

The boundary conditions are set to

(10) $$\begin{cases} \eta\left(x = w(V_l)\right) = 1 \\ \left.\frac{d\eta}{dx}\right|_{x=D} = 0 \end{cases}$$

where the first (Dirichlet) condition derives from the continuity of $\eta(x; V_l)$ at the boundary between the depletion and neutral region (x=w($V_l$)) and the second (homogeneous Neumann) condition is justified by the presence of a back surface field at the interface between the high and low doped n-type region (at x=D), which introduces a barrier to minority carrier flow to the rear [7]. In summary

(11) $$\eta(x; V_l) = \Theta[x < w(V_l)]\cdot\Theta[x > t + d] + \Theta[x < D]\cdot\Theta[x > w(V_l)]\cdot\frac{\cosh\left(\frac{D-x}{L_h}\right)}{\cosh\left(\frac{D-w(V_l)}{L_h}\right)}$$

where $\Theta(x)$ is the Heaviside step function and the last term is the solution of eq. (9), with the boundary conditions (10), assuming constant $L_h$. A detailed derivation of eq. (11) is given in the Appendix.

For each applied bias voltage $V_l$, the best fit of the experimental data through eqs. (8) and (11) has been calculated by scanning the residual (i.e. square of deviations of the theoretical curve from the experimental points) sum of squares, at constant applied bias voltage $V_l$, while varying the free parameters $w(V_l)$ and $L_h$ (step size=0.5 µm).

The matrix $\chi^2$

(12) $$\chi^2(w(V_l), L_h) = \sum_{i=1}^{M}\sum_{j=1}^{N}\left[\frac{CCE_{Exper}(E_i^0, \theta_j, V_l) - CCE_{Model}(E_i^0, \theta_j, V_l; w(V_l), L_h)}{\Delta CCE(E_i^0, \theta_j, V_l)}\right]^2$$

is the sum over the tilting angles $\theta_j$ (j=1….N=5) and energies $E_i^0$ (i=1…M≤4) of the squares of the differences between the experimental data ($CCE_{Exper}$ ) and the values ($CCE_{Model}$) predicted by the model given by eqs. (8) and (11). $\Delta CCE(V; \theta_i, E_j)$ is the uncertainty deriving from the uncertainty of calibration (of the order of few %).

The best fitting parameters are then determined by the minimum of the $\chi^2(w(V_l), L_h)$ matrix. This methodology, already successfully applied to evaluate the capture coefficients of radiation induced recombination centres [18], was used because of its easy implementation in spite of the complex non-linear CCE expression in eq. 12.

The uncertainty of the measurements of the free parameters $w(V_l)$ and $L_h$ was estimated from the calculation of the inverse of the Hessian Matrix of $\chi^2$ [30]

Fig. 7 shows the plots of the $\chi^2(w(V_l), L_h)$ matrix for different applied bias voltages. For each map, it is apparent the presence of one minimum, which moves to the left, as the bias voltage decreases. Since the abscissa of the contour plots refer to the deeper boundary of the depletion layer w(V) (see the scheme of Fig.1a), this behavior refers to the monotonous increase of the w(V) function, as expected from the basic theory of the p-n junction (Fig. 8a).

A more evident connection of this result with the electrical characterization, can be extracted from the analysis of the linear behavior of w(V) vs. 1/C(V), as shown in Fig. 8b. Actually, from the standard expression for a parallel plate capacitor [7], with the spacing between the two plates given by W(V) = w(V)-(t+d) (see Fig. 1a):

(13) $$W(V) = \frac{\varepsilon \cdot A}{C(V)}$$

and from the geometrical relationship between w and W, as shown in Fig. 1a:

(14) $$w = \frac{\varepsilon \cdot A}{C(V)} + x_J + t$$

both the electrode effective area A (ε is the silicon dielectric constant (~ 1 pF/cm)) and the junction depth $x_j$ can be extracted from the slope and the intercept, respectively, of the linear fit of Fig. 8b.

The intercept is equal to $t+x_j$=(11.9±0.4) μm, which corresponds to a junction depth of about 5.6 μm, which is a value aligned with power semiconductor rectifier design best practice [2][7].

The effective area is (A=2.70±0.07) $mm^2$, which is in between the geometrical area of the central top electrode (1.69 $mm^2$) and the area surrounding the guard rings (2.89 $mm^2$).

Finally, the minority carrier (hole) lifetime $L_h$ can be calculated from the vertical position of minima, which persist in the interval 20-30 μm, with an average value of $L_h$= (24.0±0.3) μm; assuming a hole diffusivity in silicon of 13 $cm^2/s$ [7], the hole lifetime is about 0.4 μs.

In Fig. 6 the solid lines are the plot of the $CCE_{model}$ function calculated through eq.(8), using the parameters w(V) and $L_h$ resulting from the best-fit procedure. It is remarkable the excellent agreement of the model output with the experimental data.

The η(x) profiles at different bias voltages resulting from eq. (11), are plotted in Fig. 5 and shed light on the interpretation of the complex behavior of CCE data at low bias voltages shown in Fig. 6.

As already mentioned, for bias voltages larger than 50 V, all the generation profiles are confined within the depletion layer and the decreasing CCE behaviors are caused by the increasing energy losses in the Al electrode and dead layer.

At 20 V, for proton energies of 1700 and 1500 keV, as for higher voltages, all the generation profiles are confined within the depletion layer and the decreasing CCE behaviors are caused by the increasing energy losses in the Al electrode and dead layer. However, for an energy of 2000 keV at angles smaller than 30°, the non-negligible fraction of carriers generated outside the depletion region partially contribute to the induced signal, since the time scale of carrier diffusion in the neutral region is much longer than the drift time, and, consequently, carriers are subjected to recombination phenomena. This interpretation can then be extended to lower voltages.

A similar analysis was carried out on the type 1 diode; Fig. 9 shows the CCE behaviors as a function of the incident angle, at a constant applied bias voltage, for different ion energies.

Unlikely type 2 diode, the CCE curves of diode 1 systematically decrease monotonically as a function of the tilting angle, regardless the ion energy or bias voltage. To interpret these data, we applied the same data analysis methodology adopted for type 2 diode; the fitting curves provide an excellent agreement with the experimental data, as shown by the solid lines in Fig. 9.

The best fit procedure provides w(V) relationship very similar to that shown in Fig. 8, which evidences a negligible contribution of lifetime killers to the electrostatics of the diode.

On the other hand, the minority carrier diffusion length is remarkably higher, $L_h$ = (71±4) μm. This is clearly shown in Fig. 10, where the CCE profiles η(x) show a logarithmic-like slope in the neutral region smaller than type 2 diode (see Fig. 5); the contribution of the minority carriers generated in the neutral region to the total induced charge is higher than 80% through all the whole generation profiles shown in Fig. 10 even for low bias voltages. Main conclusion is that there is a not significant loss in induced charge due to the recombination of minority carriers in the neutral region as is evidenced by the flatness of the CCE curves relevant to 2 MeV proton energy at low bias voltage. As above mentioned, the decreasing of the CCE curves, more evident at high bias voltages, is due to the energy lost along the path crossing the electrode and dead layer, whose length increases as the tilting angle increases.

Finally, the hole diffusion length evaluated for the type 1 diode has to be considered as an indicative value. Assuming a hole diffusivity in silicon of 13 $cm^2/s$ [7], the corresponding hole lifetime is about 4 μs and the average diffusion time is about few microseconds. Since the shaping time of the amplifier is 2 μs, we cannot exclude that ballistic deficit [25] can effectively affect the pulse heights, overall at low bias voltages. Therefore, the resulting $L_h$ value is more correctly to be considered as a minimum threshold value for type 1 diode.

## 5. Conclusions

We have presented a methodology based on the IBIC technique to characterize the electrostatics and to measure the minority carrier lifetime of power diodes. The technique relies on the use of scanning proton micro-beam to extract charge collection efficiency spectra from a sub-millimeter region of the diode. Spectra were analyzed as a function of the proton energy, ranging from 1.2 to 2.0 MeV, incident angle, from -40° to +50°, and applied bias voltage.

The experimental results were interpreted by using a simplified model based on the IBIC theory, which assumes a full induced charge collection in the depletion region, due to the very short drift time compared to the carrier lifetime, and an incomplete charge induction from carriers generated in the neutral region, caused by recombination processes which limit the injection of carriers by diffusion into the electric field region.

Data analysis was carried out by fitting the CCE values relevant to different energies vs. tilting angles at fixed bias voltages, through a consolidated methodology relying on scanning the residual (i.e. deviations of the theoretical curve from the experimental points) sum of squares while varying both the depletion layer width (w) and the minority carrier diffusion length $L_h$, which was assumed to be constant through the whole neutral region.

The best fit procedure was proven to be reliable in providing measurements of both the two free parameters for two p-i-n power diodes, providing a w(V) dependence in agreement with traditional electronic characterization and values of $L_h$, which account for the different performances of the two diodes.

The developed technique offers advantages over more traditional electrical methods or optical/ electron scanning micro-spectroscopy. Indeed, it is non-destructive and able to analyze the electronic features of semiconductor regions buried under thick electrodes, hardly accessible with electron or optical probes.

The methodology was inspired by work carried out by A. Lo Giudice et al. [22] regarding the angle resolved IBIC analysis of a 4H SiC Schottky diode. However, with respect to that approach, which used a monochromatic ion probe, the methodology described in this paper offers several advantages:

a) the combination of different angles of ion beam incidence and applied bias voltages, allows the diode epitaxial region to be probed. However, the use of ion probes with different energies not only provides a remarkable increase of statistics of the experimental data, but extends the probed depth from the junction to a depth of more than 40 μm from the silicon surface, with generation profiles, which, altogether, cover almost uniformly the investigated region
b) This procedure allows a consistent calibration of the electronic chain using all the ion energies, without any reference detector and is particularly suited for devices with thick electrodes, where the only basic linear relationship between pulse height and ion energy cannot be adopted, due to the not negligible energy loss in the electrode and dead layer.
c) Finally, the overall statistical data analysis, borrowed from [18], allows the measurement of the minority carrier lifetime in the intrinsic region, and the dependence of the depletion layer extension as a function of the bias voltage, which is related to the output of electrical characterizations.

Moreover, the adopted simplified model relies on hypotheses (e.g. constant carrier lifetime in the neutral region), which are reasonably acceptable for power devices and provides material parameters, which are fundamental for the calibration of device simulations with Technology Computer Aided Design (TCAD).

## 6. Acknowledgements

This work was supported by the grant "Departments of Excellence" (L.232/20165), funded by the Italian Ministry of Education, University and Research (MIUR) and by the Italian Institute for Nuclear

Physics (INFN) within the experiment “ASIDI”. Experiments were carried out with support of the RADIATE project under the Grant Agreement 824096 from the EU Research and Innovation programme HORIZON 2020.

## 7. Appendix – Derivation of eq. (11)

From the model formulated in [18], the 1D adjoint form of the continuity equations for carriers in semiconductors are

[A1] $$\begin{cases} -v_e \cdot \dfrac{\partial n^+}{\partial x} + \dfrac{\partial}{\partial x}\left[D_e \dfrac{\partial n^+}{\partial x}\right] - \dfrac{n^+}{\tau_e} + G_e^+ = 0 \\ +v_h \cdot \dfrac{\partial p^+}{\partial x} + \dfrac{\partial}{\partial x}\left[D_h \dfrac{\partial p^+}{\partial x}\right] - \dfrac{p^+}{\tau_h} + G_h^+ = 0 \end{cases}$$

Where $D_{e,h}$,$v_{e,h}$, $\tau_{e,h}$ are the diffusion coefficient, drift velocity and carrier lifetime, for electrons (e) and holes (h), respectively. $n^+$ and $p^+$ are the adjoint carrier concentrations of electrons and holes, respectively. The adjoint source terms are

[A2] $$\begin{cases} G_e^+ = v_e \cdot \dfrac{\partial \mathcal{F}}{\partial V} - \dfrac{\partial}{\partial x}\left[D_e \dfrac{\partial \mathcal{F}}{\partial x}\right] \\ G_h^+ = v_h \cdot \dfrac{\partial \mathcal{F}}{\partial V} + \dfrac{\partial}{\partial x}\left[D_h \dfrac{\partial \mathcal{F}}{\partial x}\right] \end{cases}$$

where $\frac{\partial \mathcal{F}}{\partial V}$ is the Gunn’s field, $\mathcal{F}$ the electric field and V the applied potential.

The solutions of eq. [A1] provide the charge Q(x) collected at the electrode and induced by a point charge generated at x:

[A3] $$Q(x) = q \cdot [n^+(x) + p^+(x)] = q \cdot \eta(x)$$

Where $\eta(x)$ is the spatial profile of the charge collection efficiency defined in section 0.

The total charge induced at the sensing electrode by the motion of EHPs generated with an arbitrary profile B(x) is then given by

[A4] $$Q_{Induced} = \int_0^D dx[Q(x) \cdot B(x)]$$

In the case of study, the generation profile is

[A5] $$B(x) = \frac{1}{\varepsilon_{eh}} \frac{dE_{ion}}{dx}$$

Where $E_{ion}$is the ion energy.

Eq. (8) is easily derived by the normalization of [A4] by the total charge (electrons or holes) generated by the ion:

[A6] $$Q_{Tot} = q \frac{E_{ion}}{\varepsilon_{eh}}$$

For the case of study, we assume that $n^+(x)$ and $p^+(x)$ are null when the charge is generated within the electrode and the silicon dead layer, i.e.

[A7] $\eta(x) = 0$ for x<t+d.

The rest of the integration domain of eq. [A1], i.e. $x \in [t + d, D]$ , is divided in two regions, similarly to the elementary approach adopted for the abrupt junction basic model [7]: the depletion region, extending in the range (t+d<x<w) and the neutral region extending in the range (w<x<D).

In the former region, a strong electric field occurs and the Gunn's term is $\frac{\partial \mathcal{F}}{\partial V} = \frac{1}{W}\Theta[x > t + d] \cdot \Theta[x < w]$, where W=w-(t+d) is the width of the depletion region (see Fig. 1a).

We assume that, by virtue of the high electric field in this region, the diffusion terms are negligible, and eq. [A1] can be simplified as follows

[A8] $$\begin{cases} v_e \cdot \frac{\partial n^+}{\partial x} - \frac{n^+}{\tau_e} = -\frac{v_e}{W} \\ v_h \cdot \frac{\partial p^+}{\partial x} + \frac{p^+}{\tau_h} = +\frac{v_h}{W} \end{cases}$$

Whose solutions are [18]:

[A9] $$\begin{cases} n^+(x) = \frac{1}{W}\int_x^w dy \left\{\exp\left[-\int_x^y \frac{dz}{l_e(z)}\right]\right\} \\ p^+(x) = \frac{1}{W}\int_{t+d}^x dy \left\{\exp\left[-\int_y^x \frac{dz}{l_h(z)}\right]\right\} \end{cases}$$

Finally, assuming that the drift time much shorter than the carrier lifetime through the whole depletion region, or the drift length ($l_{e,h} = v_{e,h} \cdot \tau_{e,h}$) is much shorter than W, the two functions in eq. [A9] can be simplified as

[A10] $$\begin{cases} n^+(x) = 1 - \frac{x-(t+d)}{W} \\ p^+(x) = \frac{x-(t+d)}{W} \end{cases}$$

It follows from eq. [A3] that

[A11] $\eta(x)=n^+(x)+p^+(x)=1$ in the depletion region (i.e. t+d<x<w).

For x>w, we assume that the electric field $\mathcal{F}$ is null for $x \in [w, D[$ , but not necessarily null at the back boundary, i.e. for x≅D; however, the electrostatics in this region is assumed not to be influenced by the external potential. Therefore, the Gunn's term $\left(\frac{\partial \mathcal{F}}{\partial V}\right)$ and then the adjoint source terms ($G_{e,h}$) are null. The eqs. [A1] can be then written as follows:

[A12] $$\begin{cases} -\frac{q}{k_B \cdot T} \cdot \mathcal{F}(x) \cdot \frac{\partial n^+}{\partial x} + \frac{\partial^2 n^+}{\partial x^2} - \frac{n^+}{L_e^2} = 0 \\ +\frac{q}{k_B \cdot T} \cdot \mathcal{F}(x) \cdot \frac{\partial p^+}{\partial x} + \frac{\partial^2 p^+}{\partial x^2} - \frac{p^+}{L_h^2} = 0 \end{cases}$$

where the Einstein relationship [7] has been used: $\frac{q}{k_B \cdot T} \cdot \mathcal{F} = \frac{v_{e,h}}{D_{e,h}}$ ($k_B$ is the Boltzmann constant ant T is the temperature), and $L_{e,h} = \sqrt{D_{e,h} \cdot \tau_{e,h}}$ is the carrier diffusion length.

The continuity of the solutions [A10] at the boundary at x=w, gives the Dirichlet's boundary conditions:

[A13] $$\begin{cases} n^+(x = w) = 0 \\ p^+(x = w) = 1 \end{cases}$$

The conditions at x=D depends on the diode structure. If the back contact is ohmic [31]

[A14] $$\begin{cases} n^+(x = D) = 0 \\ p^+(x = D) = 0 \end{cases}$$

and the electric field is null ($\mathcal{F}(x = D) = 0$). Therefore, the solutions of eqs. [A12] with the boundary conditions [A13] and [A14] are

[A15] $$\begin{cases} n^+(x) = 0 \\ p^+(x) = \dfrac{\sinh\left(\frac{D-x}{L_h}\right)}{\sinh\left(\frac{D-w)}{L_h}\right)} \end{cases}$$

as given in [22]. It is worth noticing that for (D-w)>> $L_h$, $p^+(x) \cong \exp\left(-\frac{x-w}{L_h}\right)$, which is the solution usually adopted in the diffusion-drift model [16].

In the case of study, the solution $p^+$ given by [A15] proved to be unsuitable to fit the experimental data. Actually, the presence of the concentration gradient at x=D, i.e. at the interface between the intrinsic region and the highly doped substrate, generates a counter electric field, oriented towards the depletion layer, operating as a reflector for the minority carrier (holes) and as a drain for the electrons.

Therefore, the boundary condition for electron is $n^+(x=D)=0$ because of the presence of an electric field, which drains the electrons into the external circuit and the solution of the first differential equation [A12], is

[A16] $n^+(x) = 0$ for w<x<D.

For the holes, we assume that the back surface field as an impenetrable barrier at x=D, i.e.$\mathcal{F}(x)$ tends to infinity for x=D. This implies that, to avoid divergences,

[A17] $$\left.\frac{\partial p^+}{\partial x}\right|_{x=D} = 0$$

which means an ideal total reflection of holes at x=D.

The hole term $p^+$ is then the solution of the diffusion equation

[A18] $$\frac{\partial^2 p^+}{\partial x^2} - \frac{p^+}{L_h^2} = 0$$

With the homogeneous Neumann boundary condition [A17] at x=D, and Dirichlet boundary condition at x=w, i.e.

[A18] $$\begin{cases} p^+(x = w) = 1 \\ \left.\frac{\partial p^+}{\partial x}\right|_{x=D} = 0 \end{cases}$$

The solution of eq. [A18] with the boundary conditions [A19] is then given by

[A20] $$p^+(x) = \frac{\cosh\left(\frac{D-x}{L_h}\right)}{\cosh\left(\frac{D-w}{L_h}\right)} \text{ for } w<x<D$$

The solutions [A7,A11, A16, A20] demonstrates the validity of eq. (11).

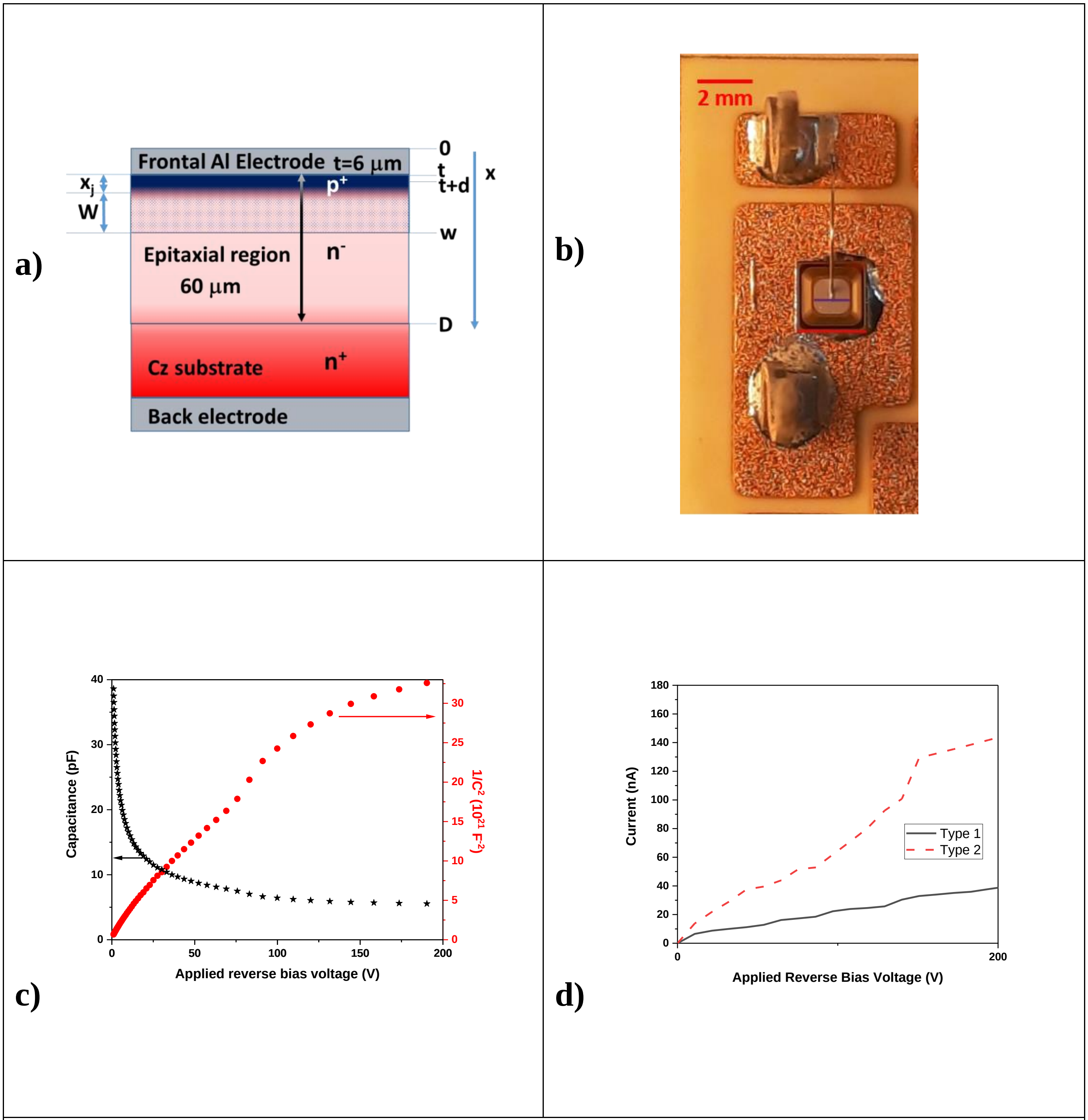

Fig. 1:a) scheme of the diode structure with the orientation of the x-axis and the main parameters (t,d,w,D); the origin of the axis is located at the electrode surface; xj is the junction depth calculated from the silicon surface and W is the depletion layer width. b) image of one diode mounted on the PCB; c) Capacitance-Voltage characteristic (left scale) and $1/C^2$ vs. V (right scale) of both the diodes; d) Current-Voltage characteristics in the reverse bias condition for the two diodes,

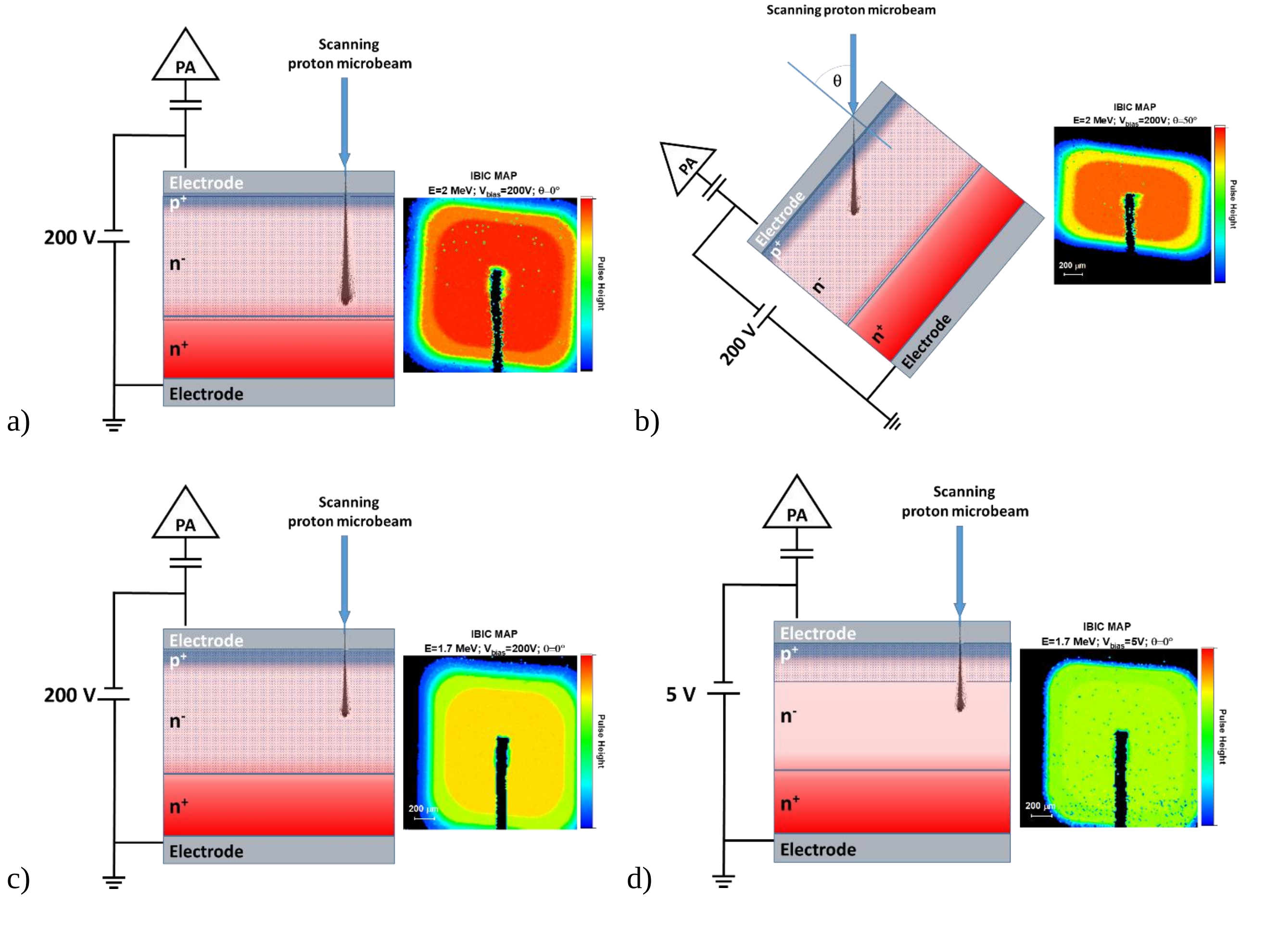

Fig. 2: Scheme and IBIC maps for different points of the parameter space (E,θ,V): a) (2 MeV, 0°, 200 V), b) (2 MeV, 50°, 200 V), c) (1.7 MeV, 0°, 200 V), d) (1.7 MeV, 0°, 5 V)

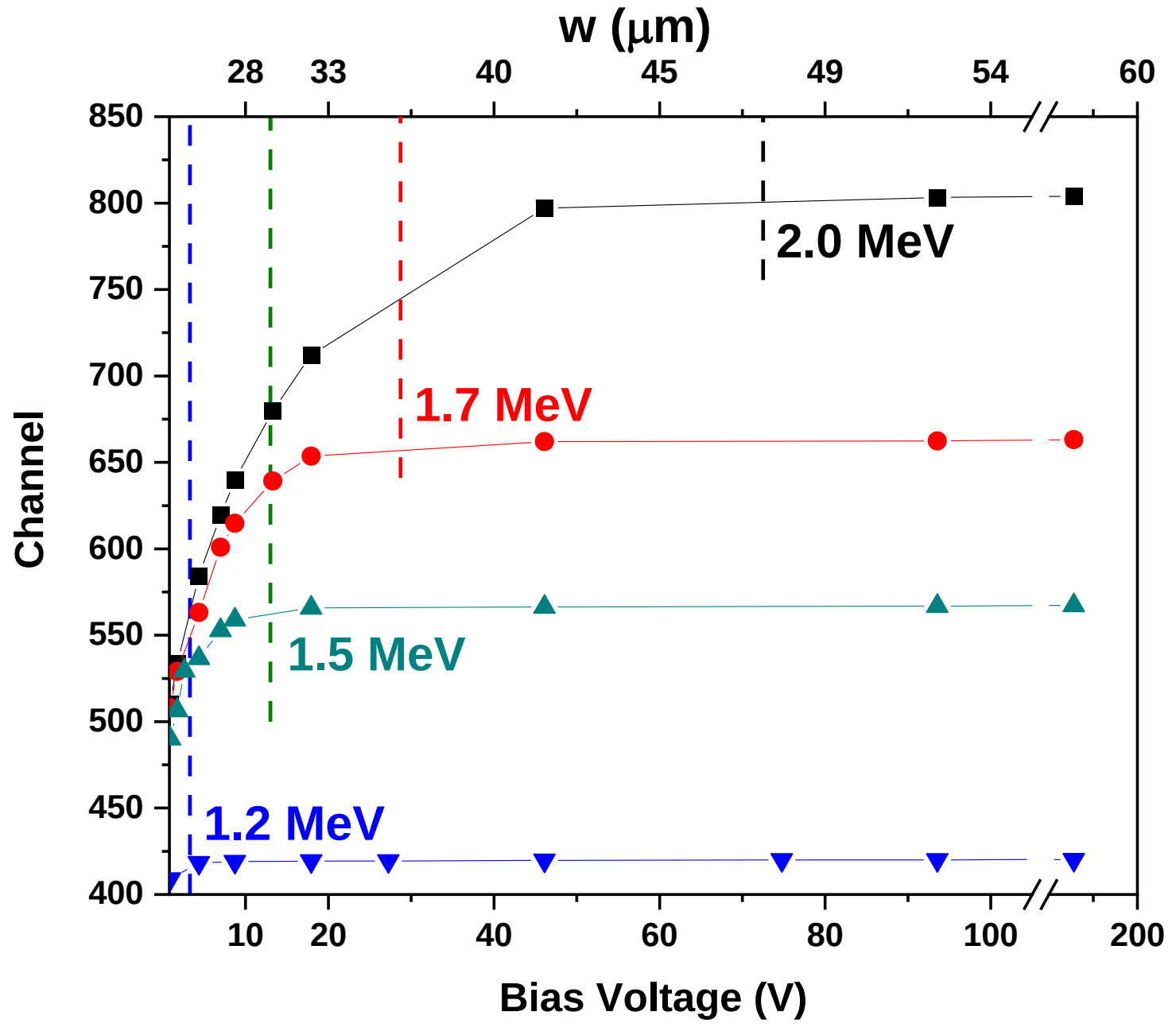

Fig. 3: pulse heights as a function of the applied bias voltage (bottom abscissa) for different proton energies. Top abscissa: depth of the depletion layer (w, refer to Fig.1), calculated through eqs. 0 and 0. Vertical lines indicate the proton range (refer to the top abscissa) in the Al+Si diode structure for different energies

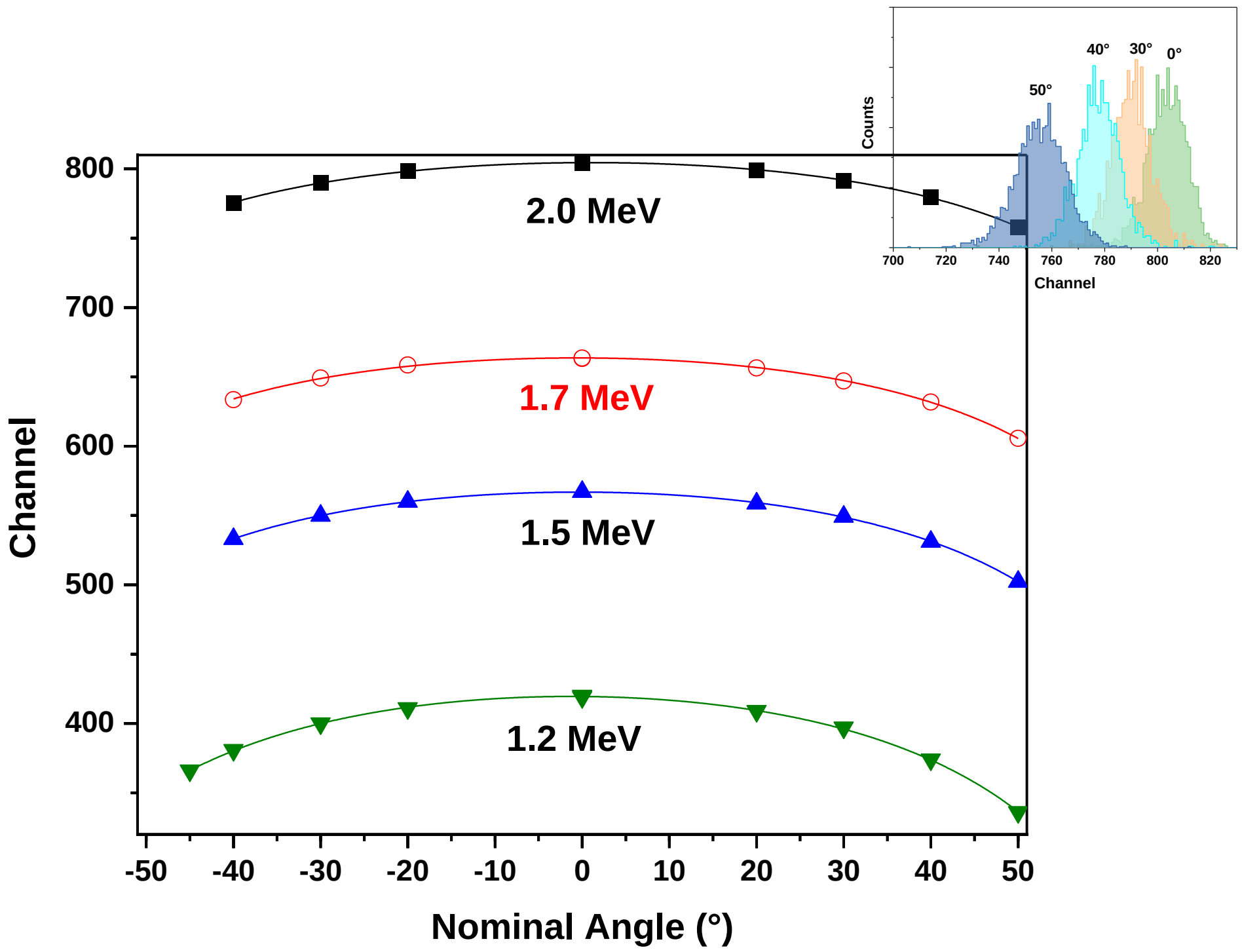

a)

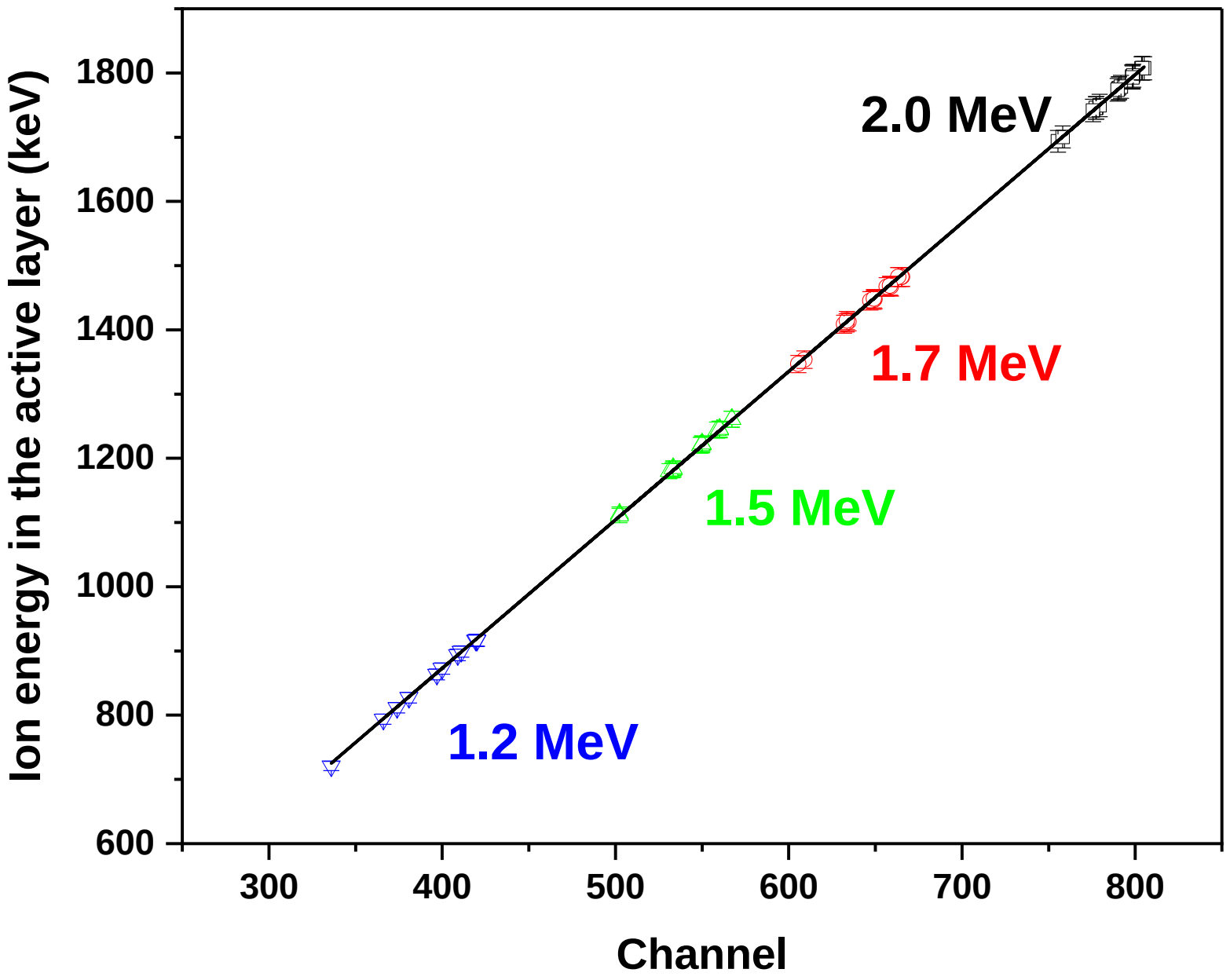

b)

Fig. 4: a) Mean pulse heights as a function of the tilting angle for different proton energies (type 2 diode). Applied bias = 200 V. Solid lines are fit of the data through eq. 5. Inset: IBIC spectra at different tilting angle. Ion probe: 2 MeV $H^+$; b) output of the fitting procedure detailed in the text; markers refer to both the diodes

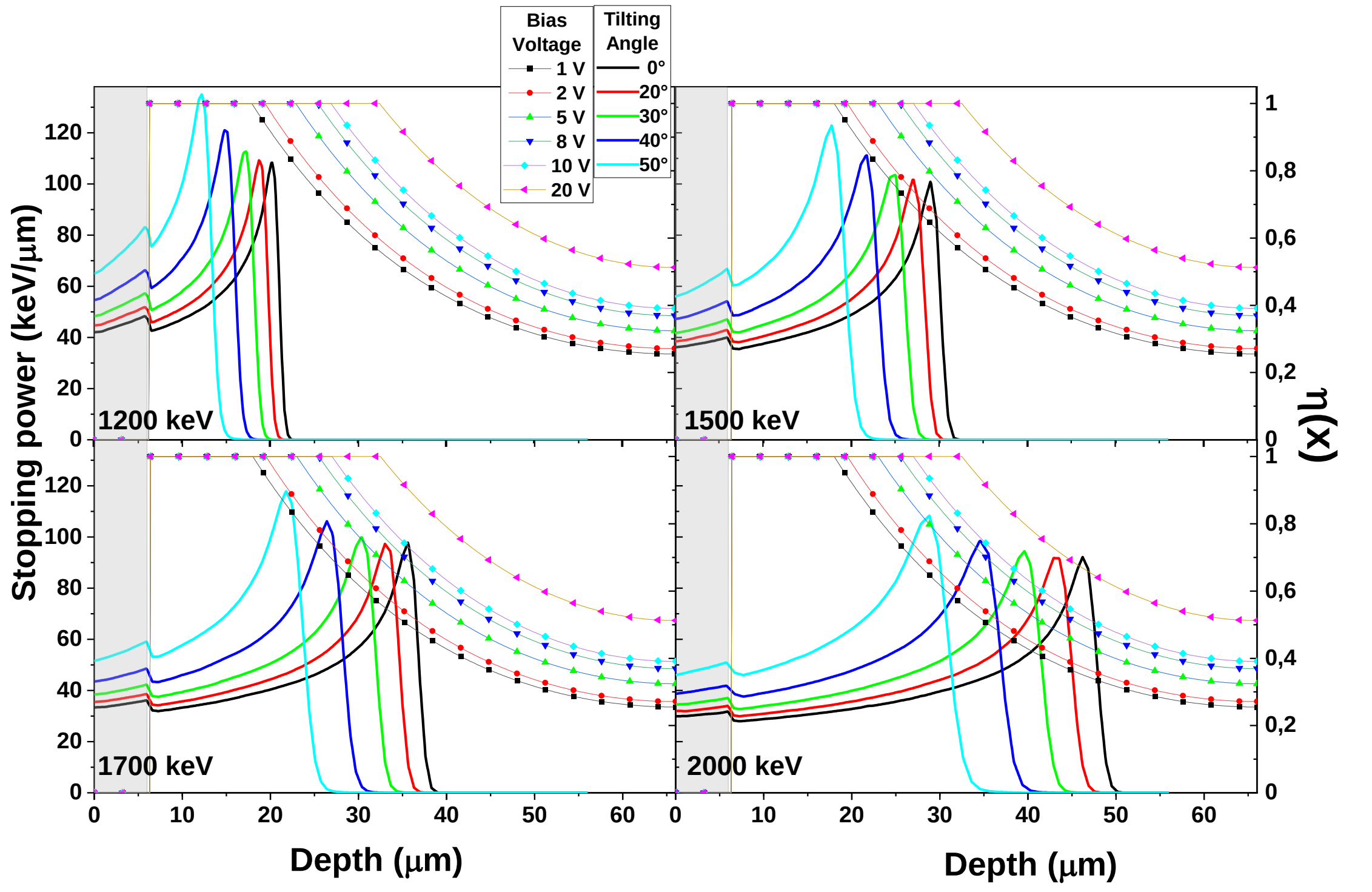

Fig. 5: solid lines, left scale: stopping power of protons at different energies and different angle of incidence in the diode, evaluated by SRIM simulations [17]. Lines + symbol graphs: CCE profile η(x) for type 2 diode for different bias voltages. The grey rectangles represent the frontal Al electrode.

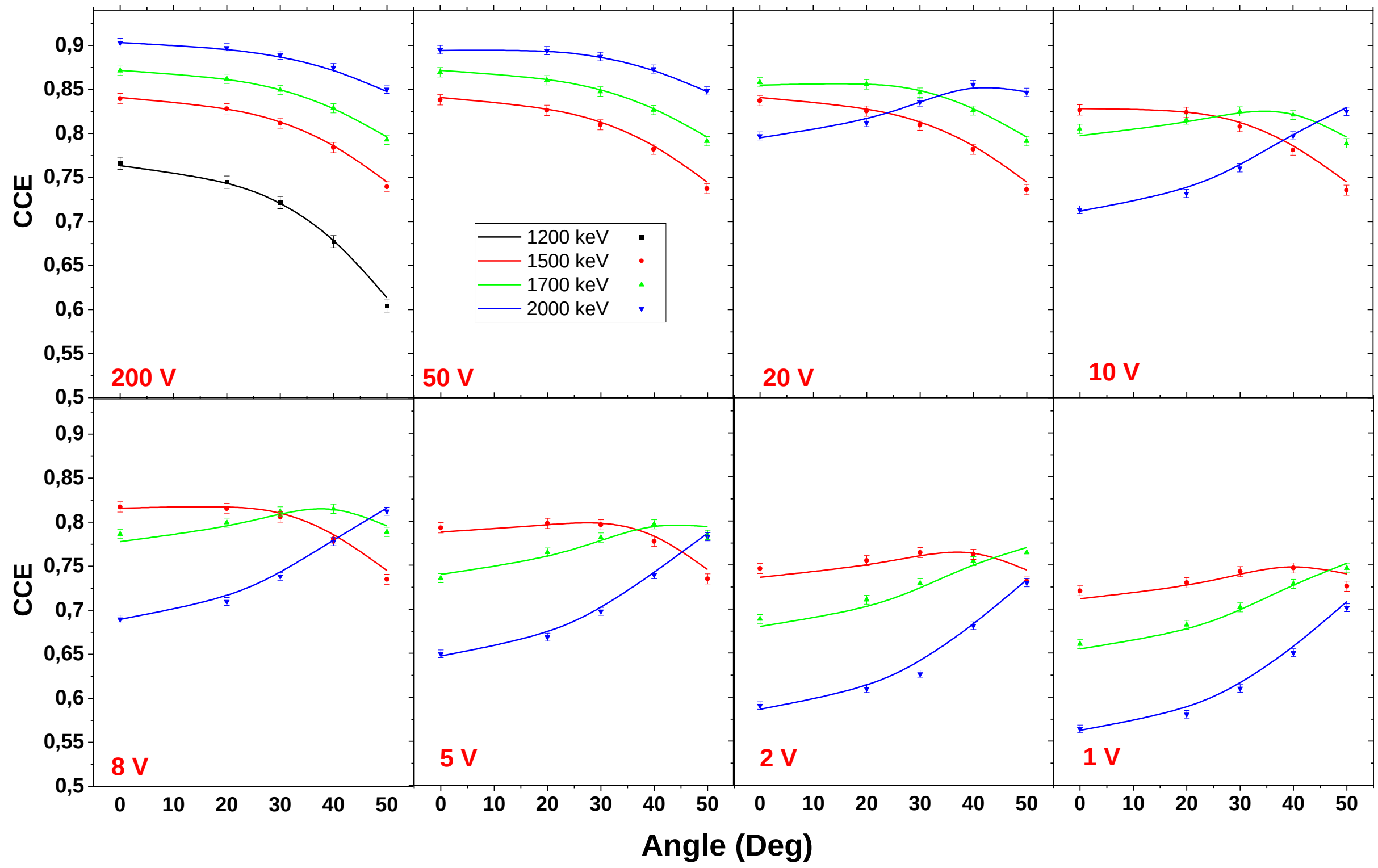

Fig. 6: CCE spectra as a function of the incidence angle parametrized by the ion energies, for different bias voltages. Solid lines are not interpolations of the experimental data (markers), but the results of the fitting procedure. Error bars result from the propagation of the uncertainty derived from the calibration procedure. Type 2 diode.

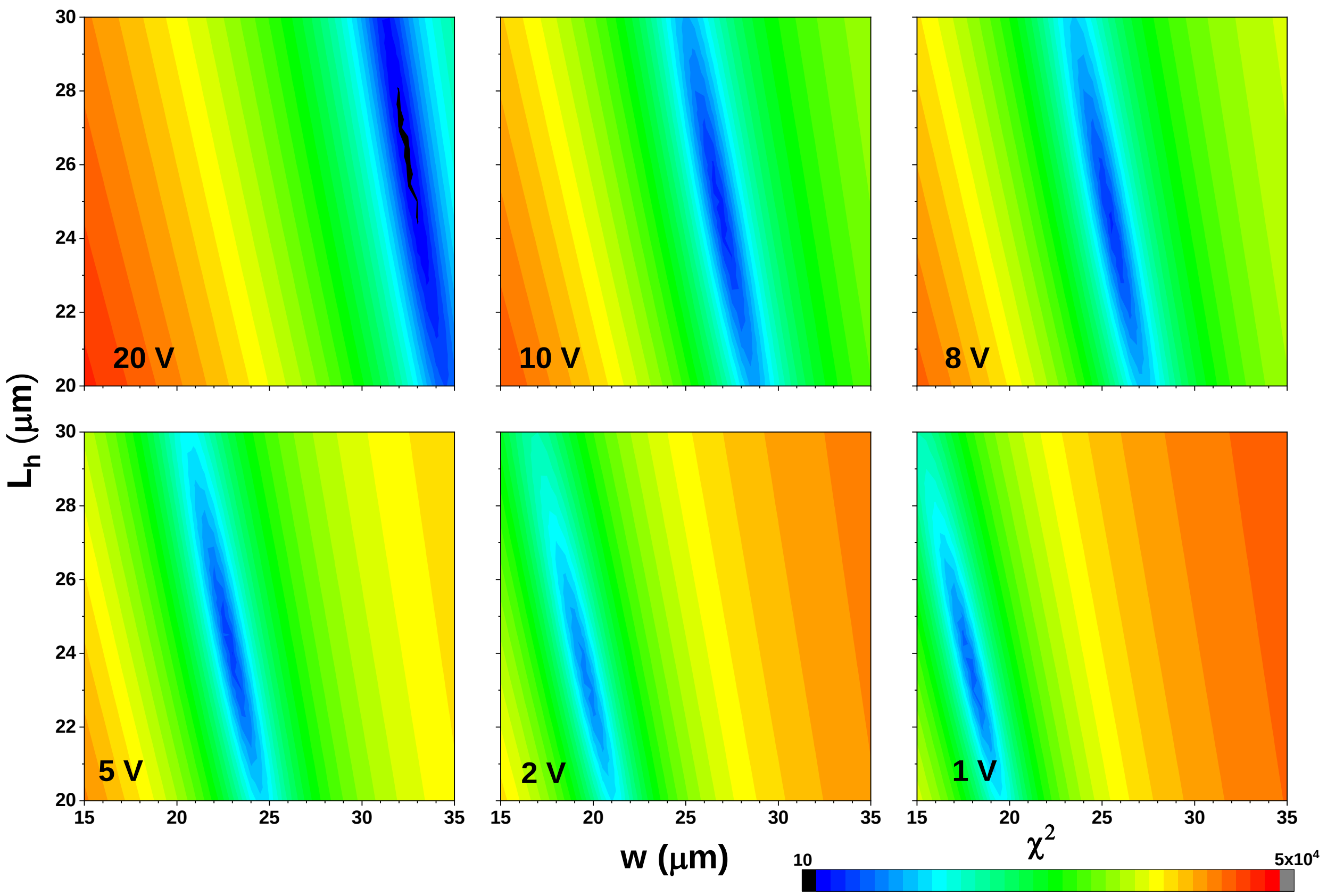

Fig. 7: . Plots of the squared sum of residuals ($\chi^2$) as a function of different values of the depletion layer width (w) and of the hole diffusion length ($L_h$) at different bias voltages. Type 2 diode.

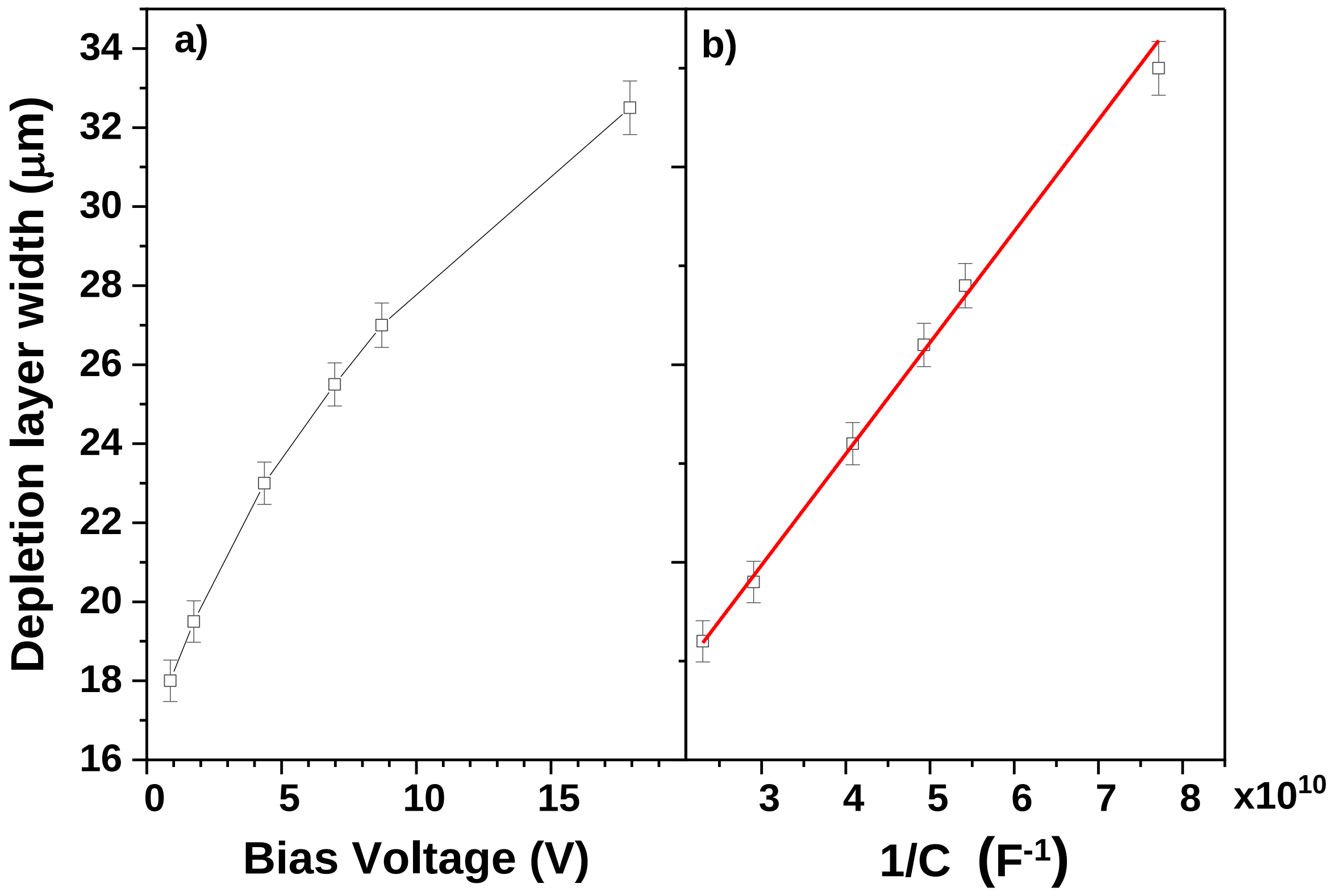

Fig. 8: depletion layer width resulting from the fitting procedure as a function of the applied bias voltage a) and of the inverse of the diode capacitance b). Type 2 diode.

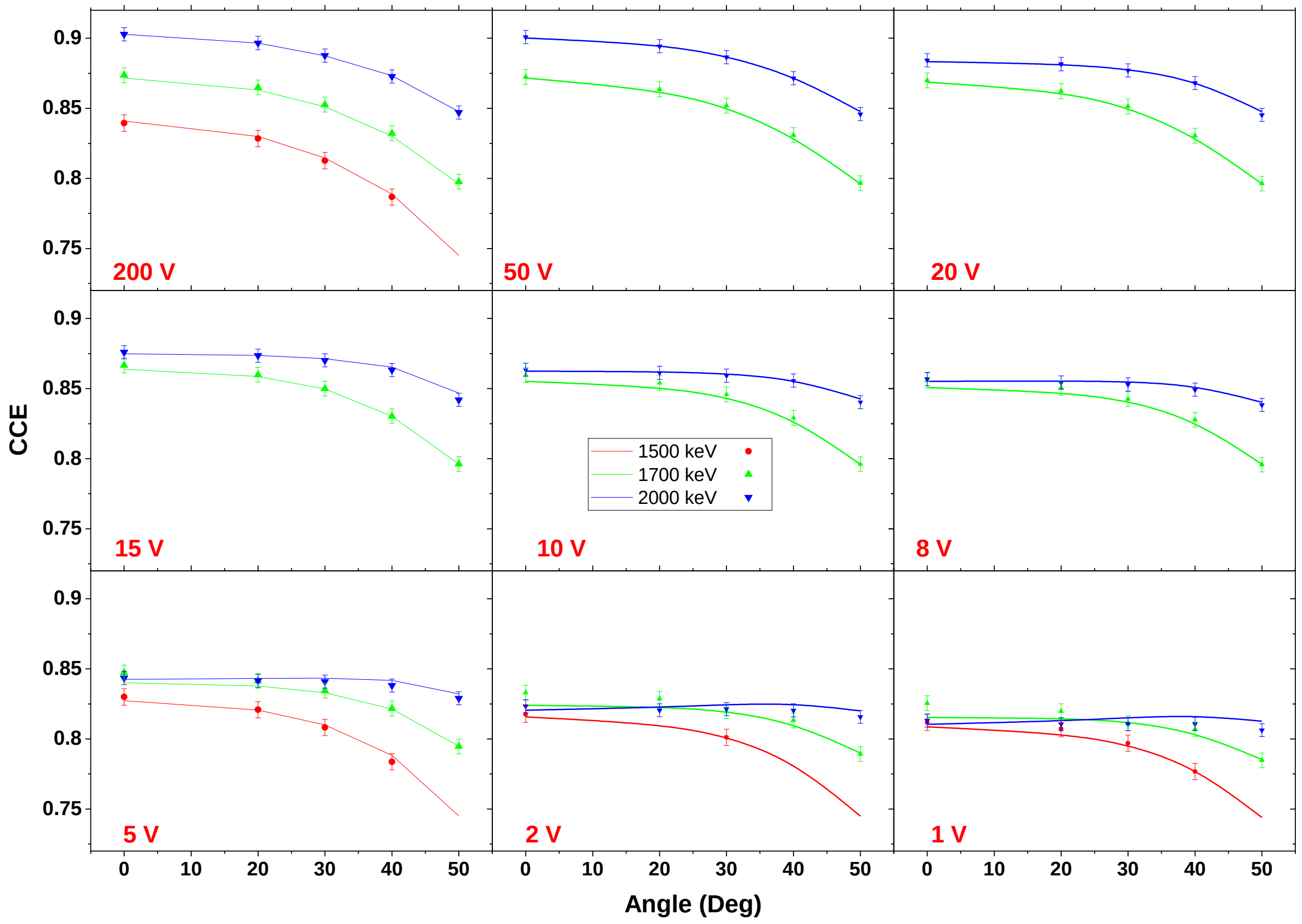

Fig. 9: same as in Fig. 6. Type 1 diode.

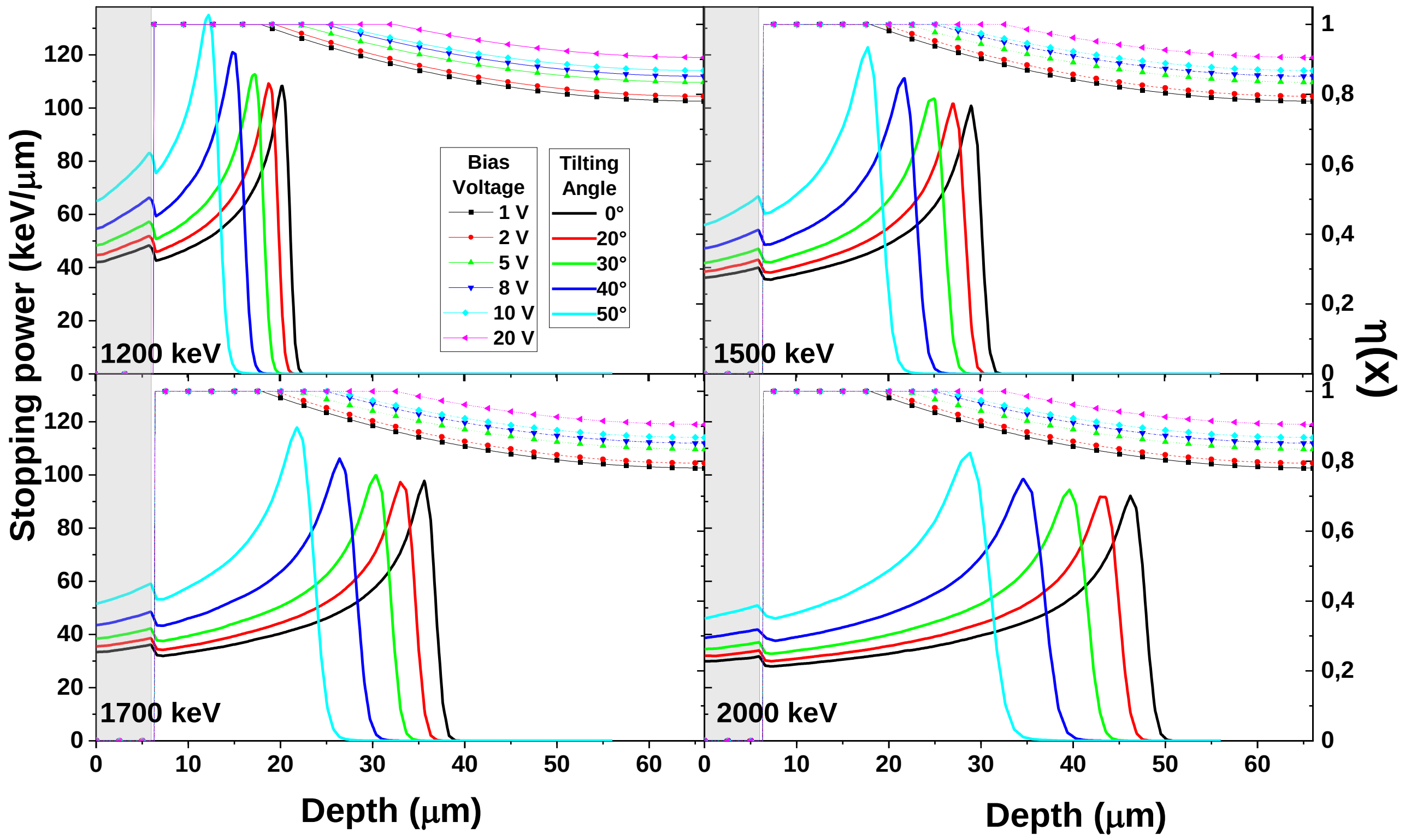

Fig. 10: same as in Fig. 5, but the CCE profile η(x) refers to type 1 diode.